# A new closed-form model for isotropic elastic sphere including new solutions for the free vibrations problem

E Hanukah [†]

[†]Faculty of Mechanical Engineering, Technion – Israel Institute of Technology, Haifa 32000, Israel
email: eliezerh@tx.technion.ac.il

**Abstract**

We develop a new closed-form model for the dynamics of an elastic (Saint Venant-Kirchhoff) and isotropic sphere. Systematic kinematic approximation by means of Taylor multivariate expansion with respect to Cartesian coordinates, and use of spherical coordinates which impose trigonometric terms to simple algebraic shape functions, is used. Constitutive relations for internal forces follow from weak (Bubnov-Galerkin) formulation together with analytical integration. Straightforward linearization of the system with respect to internal degrees of freedom leads to mass and stiffness matrices, such that small vibration problem is equivalent to algebraic eigenvalue problem. Simple expressions for resonant frequencies (eigenvalues) in terms of geometric and material constants are derived. In particular, the problem of small vibrations about the initial configuration is considered. Our second order closed form solution significantly improves currently available analytical solutions derived by Pseudo Rigid Body model. The improvement is not only in accuracy, but also in the ability to describe the response spectrum and to capture the lowest frequencies. Our third order closed form solution improves the second order solution in accuracy and the ability to represent additional modes and frequencies from the response spectrum. In addition, we derive, for the first time, a simple explicit expression for the fundamental (lowest) frequency covering the entire range of practical engineering materials (positive Poison's ratio) with high accuracy (0.02% error compared to FE results) based on a fourth order approximation.

## 1. Introduction

Development of dynamic models for elastic spheres is of practical interest (see for example[1] chapter 8). Studying the vibrations of elastic three dimensional bodies is a well-known problem as well. In the present study new straightforward formulation is developed and applied to free vibration problem resulting in new closed-form solutions for natural frequencies.

In the last several decades there has been an effort to provide 3-D elasticity solutions for the free vibration of prisms, parallelepipeds, cylinders etc., vast majority of these works have utilized various techniques to obtain *numerical results* (e.g. [2-16]).

Only a few studies applied structural theories that provide *closed-form* solutions for 3D elasticity problems (e.g. [17, 18]). These models take advantage of particular constitutive relations, justified by simplified assumptions on the structural behavior, that lead to a closed form equations of motion in terms of the degrees of freedom and also material and geometrical constants. The structural approaches which can be applied to 3-D problems can be roughly divided to two main classes: Pseudo Rigid Body and Cosserat Point, e.g. [19-23]. Cosserat Point method defines kinematic approximations in terms of directors, and also imposes restrictions on the strain energy function. In particular, the strain energy is separated into two parts; one uses average measures of deformation and the other is restricted to admit exact solutions, such as simple shear, simple torsion and pure bending, in an average sense. Cosserat point analytically satisfies Patch test, which ensures convergence when structure dimensions (size) tend to zero, see for example [24-27]. Pseudo-Rigid body is a Cosserat-like approach that enables closed form model for 3-D elastic solids of finite size. In addition to rigid body motion, it allows homogeneous deformation. Papadopoulos [28] has developed a second order theory of a pseudo-rigid body which has 30 degrees of freedom. It seems that the establishment of a pseudo-rigid body model on the basis of continuum mechanics is a delicate and unresolved issue (see [22, 29, 30]). The pseudo-rigid body method is used to simulate dynamics of multi-body systems which combine deformable solids with objects modeled as rigid bodies, e.g. [18, 31-36].

For the best of our knowledge, the only study that provides closed form expressions for the natural frequencies of an isotropic elastic sphere is [18], which builds on a Pseudo Rigid Body theory. Unfortunately, as it is revealed by present study, these frequencies (which match homogeneous deformation modes) are not the fundamental ones, lower frequencies exist.

The present study follows basic guidelines in [37] together with convenient coordinate transformation, to formulate the governing equations of motion for a 3-D sphere. In particular, the 3-D elasticity problem is converted to a set of closed form non-linear ODEs. However, a kinematic approximation is systematically derived by means of Taylor's multivariable expansion with respect to Cartesian coordinates. Spherical symmetries are not invoked, no preferable directions are assumed, which leads to general 3D kinematics. Saint Venant-Kirchhoff material model is adopted for the purpose of the study. Constitutive relations for internal forces follow from straightforward analytical integration of the weak form. The formulation is systematically expended to any desired order i.e. the dynamics of an elastic sphere can be studied using adequate accuracy level. We specialize our model to examine the free-vibrating sphere. We find that our first order solution is identical to the previously existing solution of [18], which only provides 6 nontrivial modes (matching homogeneous deformation modes) associated with *two* distinct frequencies. Our second order solution significantly improves the accuracy of the calculated frequencies. Importantly, the second order approximation predicts 24 non-trivial modes with *six* different frequencies. This is a significant improvement over the first order solution in terms of describing the response spectrum of the structure. Moreover, the solution obtained by [18] is not able to capture the lowest frequencies. Our third order solution, establishes additional accuracy improvement and additional improvement in representation of the response spectrum – *twelve* distinct frequencies (54 nontrivial modes). It has been demonstrated that different modes may be of greater engineering importance depending on Poisson's ratio, i.e. frequencies associated with different modes have a different functional dependence on Poisson's ratio. Also, we provide a highly accurate simple analytical expression for the fundamental (lowest) frequency based on a fourth order expansion.

The outline of the paper is as follows. Section 2 presents the main theoretical considerations for deriving the non-linear dynamical equations of motion. Emphasis is put on systematic derivation of a kinematic approximation in Section 2.1. Governing equations are obtained by adopting a weak formulation combined with analytical integration. Linearization of the governing equations with respect to the internal degrees of freedom is discussed in Section 3. Assuming small vibrations the mass and stiffness matrices are defined and an eigenvalue problem for the natural frequencies and modes is formulated. In Section 4, the solutions of free vibration problem are listed and discussed. The accuracy of closed-form expressions for the natural frequencies is examined by comparison to finite-elements simulations.

## 2. Theoretical considerations

In the present study we consider three dimensional sphere occupying finite volume in Euclidean space, made of isotropic St. Venant-Kirchhoff material. Governing equations of elasticity is recalled. Balance of linear momentum with respect to initial configuration (1), balance of angular momentum implies (2), St. Venant-Kirchhoff constitutive relations is given by (3), Green-Lagrange strain tensor is given by (4). Roughly speaking St. Venant-Kirchhoff is the simplest non-linear elastic constitutive model; it is a rotationally invariant model (objective or frame indifferent) such that deformations that differ by a rigid body transformation are guaranteed to have the same strain energy. Restrictions (2) imposed by balance of angular momentum are identically satisfied.

$$\rho_0 \dot{\mathbf{v}} = \rho_0 \mathbf{b} + \text{Div}(\mathbf{P}) \tag{1}$$

$$\mathbf{P}\mathbf{F}^T = \mathbf{F}\mathbf{P}^T \tag{2}$$

$$\mathbf{P} = \mathbf{F}\left(\lambda(\mathbf{E}\cdot\mathbf{I})\mathbf{I} + 2\mu\mathbf{E}\right) \tag{3}$$

$$\mathbf{E} = \frac{1}{2}\left(\mathbf{F}^T\mathbf{F} - \mathbf{I}\right) \tag{4}$$

where $\mathbf{X}$ and $\mathbf{x}$ stand for location of material point X in initial and actual configurations, so that $\mathbf{v} = \dot{\mathbf{x}}$ denotes the velocity of material particle, a superposed dot denotes time differentiation, $\rho_0$ denote mass density, $\mathbf{b}$ stands for body force per unit of mass, $\text{Div}(\bullet) = (\bullet)_{,j}\mathbf{G}^j$ denotes the divergence operator with respect to the initial configuration, $\mathbf{P}$ stand for right Piola-Kirchhoff stress tensor, $\mathbf{F} = \dfrac{\partial \mathbf{x}}{\partial \mathbf{X}}$ stand for deformation gradient, right transpose denoted by $(^T)$, $\lambda$ and $\mu$ are Lame constants. Standard relations between material constants are recalled $\lambda = E\nu/((1+\nu)(1-2\nu))$, $\mu = E/2(1+\nu)$, E is Young's modulus and $\nu$ is Poisson's ratio. $\mathbf{I}$ stands for second order identity tensor, $\Omega_0$ and $\Omega$ are initial and actual configurations (domains) of the body.

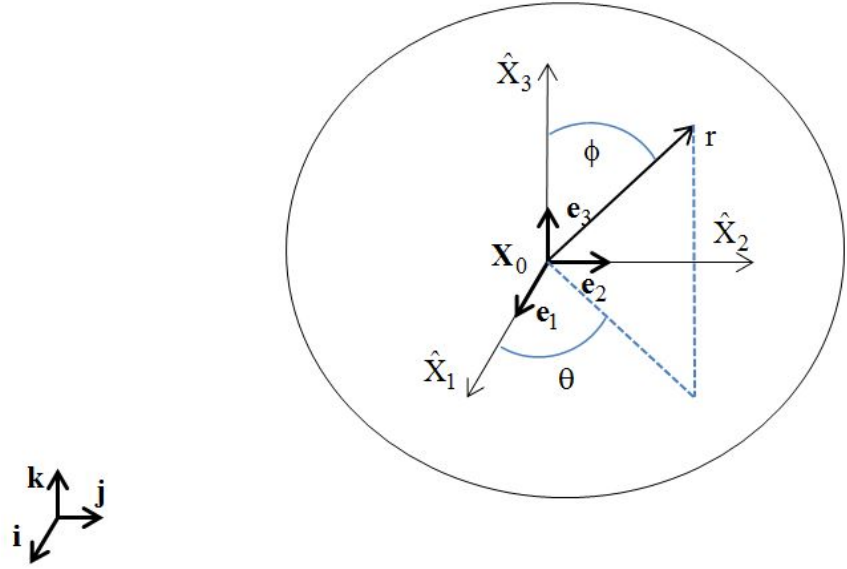

Figure 1: Schematic illustration showing the 3-D finite sphere, Cartesian orthonormal base vectors $\{\mathbf{e}_1, \mathbf{e}_2, \mathbf{e}_3\}$, their material coordinates $\{\hat{X}_1, \hat{X}_2, \hat{X}_3\}$, location of the sphere centroid $\mathbf{X}_0$, and spherical material coordinates $\{r, \phi, \theta\}$.

## 2.1 Kinematic approximation.

The initial geometry is defined by the means of centroid $\mathbf{X}_0$, three Cartesian base vectors $\{\mathbf{e}_1, \mathbf{e}_2, \mathbf{e}_3\}$; corresponding *material* coordinates $\{\hat{X}_1, \hat{X}_2, \hat{X}_3\}$ see Fig. 1. Material point $X \in \Omega_0$ occupying position $\mathbf{X}$ is exactly represented by

$$\mathbf{X} = \mathbf{X}_0 + \hat{X}_1 \mathbf{e}_1 + \hat{X}_2 \mathbf{e}_2 + \hat{X}_3 \mathbf{e}_3 \tag{5}$$

Using Lagrangian description, the actual configuration $\mathbf{x}$ of a material point $X$ is given by the motion $\mathbf{x} = \chi\left(\mathbf{X}(\hat{X}_1, \hat{X}_2, \hat{X}_3), t\right)$. The mapping function $\chi$ is unknown, and has to be determined. Finding the exact mapping is usually not possible therefore some approximations or assumptions have to be considered. Here, we systematically approximate the motion function $\chi$ by means of Taylor's multivariable expansion about the centroid of the body $\mathbf{X}_0 = \mathbf{X}(\hat{X}_1 = 0, \hat{X}_2 = 0, \hat{X}_3 = 0)$, as follows

$$\mathbf{x}^h = \chi(\mathbf{X},t)\big|_{\mathbf{X}=\mathbf{X}_0} + \sum_{k=1}^{3}\frac{\partial \chi(\mathbf{X},t)}{\partial \hat{X}_k}\bigg|_{\mathbf{X}=\mathbf{X}_0}\hat{X}_k + \sum_{k,m=1}^{3}\frac{1}{2}\frac{\partial^2 \chi(\mathbf{X},t)}{\partial \hat{X}_k \partial \hat{X}_m}\bigg|_{\mathbf{X}=\mathbf{X}_0}\hat{X}_k\hat{X}_m +$$
$$+ \sum_{k,m,n=1}^{3}\frac{1}{6}\frac{\partial^3 \chi(\mathbf{X},t)}{\partial \hat{X}_k \partial \hat{X}_m \partial \hat{X}_n}\bigg|_{\mathbf{X}=\mathbf{X}_0}\hat{X}_k\hat{X}_m\hat{X}_n + ....$$
(6)

Here and throughout the text, upper index $(\ )^h$ denotes approximated function so that $\mathbf{x}^h$ stands for approximation of $\mathbf{x}$. Partial derivatives of $\chi$ are evaluated at the centroid $\hat{X}_k = 0\,(k=1,2,3)$, thus they are all functions of time only and unknown. The order of approximation specifies the number of terms in the approximation (6) as well as the monomial terms $\hat{X}_k\hat{X}_m\hat{X}_n\,(k,m,n=0,..,3)$ which multiply the unknown (spatially constant) partial derivatives. This forms a separation of variables where all the spatial dependence is in the monomials, and all time dependence is in the derivatives. Next, it is convenient to define the shape functions $N_i\,(i=0,...,n^{shf})$

$$\begin{aligned}
&N_0 = 1 && N_1 = \hat{X}_1 && N_2 = \hat{X}_2 && N_3 = \hat{X}_3 && N_4 = \hat{X}_1\hat{X}_2 \\
&N_5 = \hat{X}_1\hat{X}_3 && N_6 = \hat{X}_2\hat{X}_3 && N_7 = \hat{X}_1\hat{X}_1 && N_8 = \hat{X}_2\hat{X}_2 && N_9 = \hat{X}_3\hat{X}_3 \\
&N_{10} = \hat{X}_1\hat{X}_2\hat{X}_3 && N_{11} = \hat{X}_1\hat{X}_1\hat{X}_2 && N_{12} = \hat{X}_1\hat{X}_1\hat{X}_3 && N_{13} = \hat{X}_2\hat{X}_2\hat{X}_1 && N_{14} = \hat{X}_2\hat{X}_2\hat{X}_3 \\
&N_{15} = \hat{X}_3\hat{X}_3\hat{X}_1 && N_{16} = \hat{X}_3\hat{X}_3\hat{X}_2 && N_{17} = \hat{X}_1\hat{X}_1\hat{X}_1 && N_{18} = \hat{X}_2\hat{X}_2\hat{X}_2 && N_{19} = \hat{X}_3\hat{X}_3\hat{X}_3\ .....
\end{aligned}$$
(7)

Using the above definitions, together with the notion that all partial derivatives in (6) can be conveniently replaced by unknown time dependent vector terms $\mathbf{x}_j(t)\,(j=0,..,n^{shf})$, the kinematic approximation (6) becomes

$$\mathbf{x}^h = \sum_{j=0}^{n^{shf}} N_j \mathbf{x}_j(t)$$

$1^{st}$ order $\Rightarrow n^{shf} = 3$ , $2^{nd}$ order $\Rightarrow n^{shf} = 9$ (8)
$3^{rd}$ order $\Rightarrow n^{shf} = 19$ , $4^{rd}$ order $\Rightarrow n^{shf} = 34$ .....

Unknown vector terms $\mathbf{x}_j(t)\,(j=0,..,n^{shf})$ have to be determined in order to estimate the deformation field. Each vector $\mathbf{x}_j(t)$ includes three components. Therefore the notion of internal degrees of freedom (IDF) arises naturally, and using (8) the total number of IDFs, $n^{dof}$, is can be expressed by simple relation

$$n^{dof} = 3\left(n^{shf} + 1\right). \tag{9}$$

Using definition (7), the exact representation (5) may be rewritten as

$$\mathbf{X} = \sum_{j=0}^{n^{shf}} N_j \mathbf{X}_j \; , \; \mathbf{X}_j = \mathbf{e}_j \; (j=1,2,3) \; , \; \mathbf{X}_j\big|_{j>3} = 0 \tag{10}$$

To determine the initial values of $\mathbf{x}_j(t)$, it is recalled that approximation (8) has to admit (10) for the initial configuration, so the initial values of unknown functions $\mathbf{x}_j(t)$ are

$$\mathbf{x}^h\big|_{\substack{t\to 0 \\ \chi\to 1}} = \mathbf{X} \;\Rightarrow\; \mathbf{x}_j\big|_{\substack{t\to 0 \\ \chi\to 1}} = \mathbf{X}_j \; , \; \left(j=0,..,n^{shf}\right) \tag{11}$$

where **1** stands for identity transformation. Next, displacement field will be approximated in the same manner, to this end it is convenient to define

$$\mathbf{u}_j(t) = \mathbf{x}_j(t) - \mathbf{X}_j \; , \; \mathbf{u}_j\big|_{\substack{t\to 0 \\ \chi\to 1}} = \mathbf{0} \; , \; \left(j=0,..,n^{shf}\right). \tag{12}$$

Thus, the approximated displacement field becomes

$$\mathbf{u}^h = \mathbf{x}^h - \mathbf{X} = \sum_{j=0}^{n^{shf}} N_j \mathbf{u}_j(t) \tag{13}$$

Since (12) implies that all $\mathbf{u}_j(t)\left(j=0,..,n^{shf}\right)$ are zero in the initial configuration, it is natural to define the internal degrees of freedom as their components, i.e.

$$\mathbf{u}_j(t) = \sum_{k=1}^{3} b_{3j+k}(t)\mathbf{e}_k \; , \; b_m(t)\big|_{\substack{t\to 0 \\ \chi\to 1}} = 0 \; , \; \left(j=0,..,n^{shf}, m=1,..,n^{dof}\right) \tag{14}$$

Standard definitions of covariant and contravariant base vectors in initial configuration and some relations between them are recalled

$$\mathbf{G}_k = \partial \mathbf{X}/\partial \hat{X}_k = \mathbf{X}_{,k} = \mathbf{e}_k \; , \; (k=1,2,3) \; , \; G^{1/2} = \mathbf{G}_1 \times \mathbf{G}_2 \cdot \mathbf{G}_3 = 1 > 0 \tag{15}$$

$$\mathbf{G}^1 = \frac{\mathbf{G}_2 \times \mathbf{G}_3}{G^{1/2}} \; , \; \mathbf{G}^2 = \frac{\mathbf{G}_3 \times \mathbf{G}_1}{G^{1/2}} \; , \; \mathbf{G}^3 = \frac{\mathbf{G}_1 \times \mathbf{G}_2}{G^{1/2}} \tag{16}$$

Where $\delta_m^n$ is the Kronecker's delta, $(\bullet)$ stands for scalar product and $(\times)$ for vector product, comma $(\bullet)_{,i}$ $(i=1,2,3)$ denotes partial differentiation with respect to $\hat{X}_k$. Next we proceed to the following standard definitions of covariant and contravariant bases and their relations

$$\mathbf{g}_k = \partial \mathbf{x}^h / \partial \hat{X}_k = \mathbf{x}^h{}_{,k} \ , \ g^{1/2} = \mathbf{g}_1 \times \mathbf{g}_2 \bullet \mathbf{g}_3 > 0$$
$$\mathbf{g}^1 = \frac{\mathbf{g}_2 \times \mathbf{g}_3}{g^{1/2}} \ , \ \mathbf{g}^2 = \frac{\mathbf{g}_3 \times \mathbf{g}_1}{g^{1/2}} \ , \ \mathbf{g}^3 = \frac{\mathbf{g}_1 \times \mathbf{g}_2}{g^{1/2}} \ , \ \mathbf{g}_k \bullet \mathbf{g}^m = \delta_k^m \ , \ (k,m=1,2,3) \quad (17)$$

where comma denotes partial differentiation. Also, using the definitions of the deformation gradient, $\mathbf{F} = \dfrac{\partial \mathbf{x}}{\partial \mathbf{X}}$, together with (17),(15),(16) and (4), the approximated forms of the kinematic tensors are

$$\mathbf{F}^h = \sum_{k=1}^{3} \mathbf{g}_k \otimes \mathbf{G}^k \ , \ \left(\mathbf{F}^h\right)^T = \sum_{k=1}^{3} \mathbf{G}^k \otimes \mathbf{g}_k \quad (18)$$

$$\mathbf{E}^h = \frac{1}{2}\left(\left(\mathbf{F}^h\right)^T \mathbf{F}^h - \mathbf{I}\right) \quad (19)$$

where $\otimes$ stands for tensor/outer product. Using the above, approximated constitutive law (3) is given by

$$\mathbf{P}^h = \mathbf{F}^h \left(\lambda \left(\mathbf{E}^h \bullet \mathbf{I}\right)\mathbf{I} + 2\mu \mathbf{E}^h\right) \quad (20)$$

where $(\bullet)$ stands for dot product between two second order tensors (double contraction). With the help of the above balance of linear momentum (1) is approximated as

$$\mathbf{R}^h = \text{Div}(\mathbf{P}^h) + \rho_0 \mathbf{b} - \rho_0 \dot{\mathbf{v}}^h \quad (21)$$

where $\dot{\mathbf{v}}^h = \ddot{\mathbf{x}}^h$. Note, generally speaking, $\mathbf{R}^h$ does not vanish identically since kinematic approximation is implied. It is a function of $n^{dof}$ internal degrees of freedom $b_p(t)\left(p=1,..,n^{dof}\right)$. Next, we adopt a weak formulation to derive $\left(n^{shf}+1\right)$ vector equations of motion and consequently $n^{dof}$ scalar ODEs.

For later integration in spherical domain, coordinates $\{\hat{X}_1, \hat{X}_2, \hat{X}_3\}$ are changed to Spherical coordinates $\{r, \phi, \theta\}$ see Fig. 1. by the next transformation

$$\hat{X}_1 = r\cos(\theta)\sin(\phi) \ , \ \hat{X}_2 = r\sin(\theta)\sin(\phi) \ , \ \hat{X}_3 = r\cos(\phi) \ , \ dV = r^2\sin(\phi)drd\phi d\theta \quad (22)$$

Where $dV$ stands for differential volume element in initial configuration.

## 2.2 Equations of motion.

A weak formulation is used to restrict the residual $\mathbf{R}^h$ in the domain, and to obtain the governing dynamical equations of motion of the system

$$\mathbf{R}_i\left(R,E,\nu,\rho_0;b_p(t),\ddot{b}_q(t)\right) = \int_{\Omega_0} \mathbf{R}^h N_i dV = \mathbf{0} \ , \ \left(i=0,..,n^{shf}\right),\left(p,q=1,..,n^{dof}\right) \quad (23)$$

Where $R$ stand for sphere radius. The above weak form defines $\left(n^{shf}+1\right)$ vector coupled ordinary differential equations. With the help of partial integration, divergence theorem and introduction of the traction boundary condition $\bar{\mathbf{t}} = \mathbf{P}\mathbf{N}$ ($\mathbf{N}$ is normal to the boundary in initial configuration), (23) becomes

$$\int_{\Omega_0} N_i \rho_0 \dot{\mathbf{v}}^h dV = \int_{\Omega_0} N_i \rho_0 \mathbf{b} dV + \int_{\partial\Omega_0} N_i \bar{\mathbf{t}} dA - \int_{\Omega_0} \mathbf{P}^h \mathrm{Grad}(N_i) dV$$
$$\left(i=0,..,n^{shf}\right) \quad (24)$$

where $\mathrm{Grad}(N_i) = \sum_{k=1}^{3} (N_i)_{,k} \mathbf{G}^k$ is the gradient operator with respect to initial configuration.

Next, definitions for mass coefficients - $\hat{M}_{ij}$, internal forces - $\mathbf{f}_i^{int}$, external and body forces - $\mathbf{f}_i^{ext}, \mathbf{f}_i^{body}$, are introduced

$$\mathbf{f}_i^{ext} = \int_{\partial\Omega_0} N_i \bar{\mathbf{t}} dA$$
$$\mathbf{f}_i^{body} = \int_{\Omega_0} N_i \rho_0 \mathbf{b} dV \ , \ (i=0,..,n^{shf}) \quad (25)$$

$$\hat{M}_{ij} = \rho_0 \int_0^R \int_0^\pi \int_0^{2\pi} (N_i N_j) \overbrace{r^2 \sin(\phi) dr d\phi d\theta}^{dV}$$

$$\mathbf{f}_i^{int} = \int_0^R \int_0^\pi \int_0^{2\pi} \mathbf{P}^h \mathrm{Grad}(N_i) \underbrace{r^2 \sin(\phi) dr d\phi d\theta}_{dV} \ , \ \left(i,j=0,..,n^{shf}\right) \quad (26)$$

With the help of (26),(13) and (8), the (vector) equations of motion (24), are written as follows

$$\mathbf{R}_i = \sum_{j=0}^{n^{shf}} \hat{M}_{ij}\ddot{\mathbf{u}}_j + \mathbf{f}_i^{int} - \mathbf{f}_i^{ext} - \mathbf{f}_i^{body} = \mathbf{0} \ , \ \left(i = 0,..,n^{shf}\right) \qquad (27)$$

Importantly, it is possible to write explicit closed-forms of equations $\mathbf{R}_i$ in terms of the initial geometry parameter $R$, and the material constants $E, \nu, \rho_0$ and scalar functions $\ddot{b}_p(t), b_q(t)$, $(p,q = 1,..,n^{dof})$. Further, integrals (26) are analytically integrated with no necessity of integration points. These symbolic evaluations are not included here, for brevity, but are obtained with the aid of commercial symbolic software Maple$^{TM}$.

## 3. Linearization and formulation of the free vibration problem

In the previous section, vector equation of motion has been derived. Herein, derivation of a set of $n^{dof}$ scalar equations of motion is presented, linearization is carried out, mass and stiffness matrices are defined, and an eigenvalue problem is formulated for the free vibration analysis. To this end, we define

$$F_{3i+k}\left(R, E, \nu, \rho_0; \ddot{b}_p(t), b_q(t)\right) = \mathbf{R}_i \cdot \mathbf{e}_k = 0$$
$$\left(i = 0,..,n^{shf}\right), (k = 1, 2, 3), \left(p, q = 1,..,n^{dof}\right) \quad (28)$$

and

$$F_{3i+k}^{mass}\left(R, \rho_0; \ddot{b}_p(t)\right) = \sum_{j=0}^{n^{shf}} \hat{M}_{ij} \ddot{\mathbf{u}}_j \cdot \mathbf{e}_k$$

$$F_{3i+k}^{stiff}\left(R, E, \nu; b_p(t)\right) = \mathbf{f}_i^{int} \cdot \mathbf{e}_k \quad (29)$$

$$\left(i = 0,..,n^{shf}\right), (k = 1, 2, 3), \left(p = 1,..,n^{dof}\right)$$

Free-vibration problem is investigated, therefore external and body forces are neglected. These enable us to compactly rewrite (28) as

$$F_p = F_p^{mass} + F_p^{stiff} = 0 \; , \; \left(p = 1,..,n^{dof}\right). \quad (30)$$

Next, we introduce the following notations

$$[\ddot{b}] = \begin{bmatrix} \ddot{b}^1 \\ \cdot \\ \cdot \\ \cdot \\ \ddot{b}^{n^{dof}} \end{bmatrix}, \; [b] = \begin{bmatrix} b^1 \\ \cdot \\ \cdot \\ \cdot \\ b^{n^{dof}} \end{bmatrix}, \; [F] = \left[F^{mass}\right] + \left[F^{stiff}\right] = \begin{bmatrix} F_1^{mass} \\ \cdot \\ \cdot \\ F_{n^{dof}}^{mass} \end{bmatrix} + \begin{bmatrix} F_1^{stiff} \\ \cdot \\ \cdot \\ F_{n^{dof}}^{stiff} \end{bmatrix} = [0] \quad (31)$$

Linearization of the system $[F] = [0]$ is carried out about the initial configuration, namely $[b] = [0]$, such that $[F] \approx [F]_{[b]=[0]} + \dfrac{\partial [F]}{\partial [b]}\bigg|_{[b]=[0]} [b]$. Using definitions(29), notation(31), the mass od stiffness matricies of the system of ODE (30) can written as:

$$[K] = \begin{bmatrix} K_{11} & . & K_{1n^{dof}} \\ . & . & . \\ K_{n^{dof}1} & . & K_{n^{dof}n^{dof}} \end{bmatrix}, \quad K_{ij} = \left. \frac{\partial F_i^{stiff}}{\partial b_j} \right|_{[b]=[0]}$$
$$[M] = \begin{bmatrix} M_{11} & . & M_{1n^{dof}} \\ . & . & . \\ M_{n^{dof}1} & . & M_{n^{dof}n^{dof}} \end{bmatrix}, \quad M_{ij} = \frac{\partial F_i^{mass}}{\partial \ddot{b}_j}$$

(32)

Both matrices are symmetric. In addition, the mass matrix is positive definite and the stiffness matrix is positive semi-definite due to rigid body motion modes. Using the above definition, the linearization of the system (30) about the initial configuration $[b]=[0]$ is given by

$$[F] \approx [M][\ddot{b}] + [K][b] = [0] \tag{33}$$

Next it is assumed that $[b] = [\tilde{b}]\sin(\omega t)$ were $\omega$ is the vibration natural frequency and $[\tilde{b}]$ is the mode of the vibration (constant algebraic vector). The second time derivative becomes $[\ddot{b}] = -\omega^2 [\tilde{b}] \sin(\omega t)$, and after substituting to (33) the well-known form for small vibration problem in many degree of freedom (MDOF) system is

$$\left(-\omega^2 [M] + [K]\right)[\tilde{b}]\sin(\omega t) = [0] \tag{34}$$

Non-trivial solutions for $[\tilde{b}]$ exists if and only if $\left|-\omega^2 [M] + [K]\right| = 0$ (determinant vanishes). Solution of this problem leads to $n^{dof}$ natural modes and frequencies. The lowest six natural frequencies are zero, and represent rigid body translation and rotation.

## 4. Natural frequencies-sphere.

It has been found that all expressions for resonant frequencies for all orders which has been addressed (1st-4st) take the next form

$$\omega_k^2 = \frac{E}{\rho_0 R^2} \bar{\omega}_k^2(\nu) \ , \ (k=1,2,3...) \tag{35}$$

where $\bar{\omega}_k$ is a non-dimensional frequency that depends only on $\nu$. This result provides an essential practical insight regarding the role of material and geometrical properties of a free vibrating sphere. Finally, we note that although some of the above expressions involve complex terms, all frequencies are real. This is a direct consequence of the fact that the mass and stiffness matrices are positive definite and semi-positive definite, respectively, in addition to both being real and symmetric.

**First order approximation.** We begin by calculating the natural frequencies associated with the first order approximation. To this end, the stiffness and mass matrices (32) are formulated and the eigenvalue problem (34) is solved. First order i.e. $\mathrm{n}^{\mathrm{dof}} = 12$, and therefore the number of natural frequencies is 12. Also, as stated earlier, the first 6 natural frequencies match rigid body motion modes (three rigid body translation and three rotation modes) are zero. The remaining 6 non-trivial natural frequencies are

$$\begin{aligned}\bar{\omega}_1^2 = \bar{\omega}_2^2 = \bar{\omega}_3^2 = \bar{\omega}_4^2 = \bar{\omega}_5^2 = \bar{\omega}^{*2} = \frac{5}{(1+\nu)} \\ \bar{\omega}_6^2 = \frac{5}{(1-2\nu)}\end{aligned} \tag{36}$$

The five frequencies, $\bar{\omega}_1$ through $\bar{\omega}_5$, are associated with two homogeneous elongation modes and three simple shear modes in the $\mathbf{e}_1 - \mathbf{e}_2, \mathbf{e}_1 - \mathbf{e}_3, \mathbf{e}_2 - \mathbf{e}_3$ planes, while $\bar{\omega}_6$ is associated with dilatational mode. The reason for the existence of only two elongation modes is that the third elongation mode is a linear combination of these two. Importantly, the results of the first order approximation, (36), are identical to those found [18]. Using first order approximation, agreement with the previously existing results is established. Next, we excel this solution by providing better accuracy and richer spectral analysis by means of higher order approximations.

**Second order approximation.** The second order approximation involves $n^{dof}=30$, namely the system has $30-6=24$ non-trivial frequencies, four times more than the previous solution (which provided 6 nontrivial expressions two of them distinct). Formulating (32) and solving the corresponding eigenvalue problem(34), we find that the first order solution - (36), is included in the second order solution.

$$\bar{\omega}_1^2 = \bar{\omega}_2^2 = \bar{\omega}_3^2 = \bar{\omega}_4^2 = \bar{\omega}_5^2 = 0.7\bar{\omega}^{*2}$$
$$\bar{\omega}_6^2 = \bar{\omega}_7^2 = \bar{\omega}_8^2 = \bar{\omega}_9^2 = \bar{\omega}_{10}^2 = \bar{\omega}^{*2}$$
$$\bar{\omega}_{11}^2 = \bar{\omega}_{12}^2 = \bar{\omega}_{13}^2 = \frac{7\left(8\nu-9+\sqrt{144\nu^2-104\nu+41}\right)}{4\left(-1+\nu+2\nu^2\right)}$$
$$\bar{\omega}_{14}^2 = \bar{\omega}_{15}^2 = \bar{\omega}_{16}^2 = \bar{\omega}_{17}^2 = \bar{\omega}_{18}^2 = \bar{\omega}_{19}^2 = \bar{\omega}_{20}^2 = 2.8\bar{\omega}^{*2} \qquad (37)$$
$$\bar{\omega}_{21}^2 = \frac{5}{(1-2\nu)}$$
$$\bar{\omega}_{22}^2 = \bar{\omega}_{23}^2 = \bar{\omega}_{24}^2 = \frac{7\left(8\nu-9-\sqrt{144\nu^2-104\nu+41}\right)}{4\left(-1+\nu+2\nu^2\right)}$$

It is immediately noted that $\bar{\omega}_7, \bar{\omega}_8, \bar{\omega}_9, \bar{\omega}_{10}, \bar{\omega}_{11}$ are smaller than $\bar{\omega}^*$. The lowest resonant frequency is called fundamental frequency, for most engineering cases it is an important one. Thus, the solution provided by the second order approximation is superior to the first order solution since it leads to better representation of fundamental frequency. Moreover, six different formulas are observed instead of two.

**Third order approximation.** The third order approximation involves $n^{dof}=60$, namely the system has $60-6=54$ non-trivial frequencies. Formulating (32) and solving the corresponding eigenvalue problem(34), we find that first order solution is not included here

$$\bar{\omega}_1^2 = \bar{\omega}_2^2 = \bar{\omega}_3^2 = \bar{\omega}_4^2 = \bar{\omega}_5^2 = 0.7\bar{\omega}^{*2}$$

$$\bar{\omega}_6^2 = \bar{\omega}_7^2 = \bar{\omega}_8^2 = \bar{\omega}_9^2 = \bar{\omega}_{10}^2 = \frac{\hat{r}^2 + 29164v^2 - 29038v + 11197 + 164\hat{r}v - 145\hat{r}}{6\hat{r}(-1+2v)(1+v)}$$

$$\bar{\omega}_{11}^2 = \bar{\omega}_{12}^2 = \bar{\omega}_{13}^2 = \frac{7\left(8v - 9 + \sqrt{144v^2 - 104v + 41}\right)}{4\left(-1 + v + 2v^2\right)}$$

$$\bar{\omega}_{14}^2 = \bar{\omega}_{15}^2 = \bar{\omega}_{16}^2 = \bar{\omega}_{17}^2 = \bar{\omega}_{18}^2 = \bar{\omega}_{19}^2 = \bar{\omega}_{20}^2 = 1.8\bar{\omega}^{*2}$$

$$\bar{\omega}_{21}^2 = \frac{49v - 77 + \sqrt{3661v^2 - 7546v + 4669}}{2(1+v)(-1+2v)}$$

$$\bar{\omega}_{22}^2 = \bar{\omega}_{23}^2 = \bar{\omega}_{24}^2 = \bar{\omega}_{25}^2 = \bar{\omega}_{26}^2 = \bar{\omega}_{27}^2 = \bar{\omega}_{28}^2 = 2.8\bar{\omega}^{*2}$$

$$\bar{\omega}_{29}^2 = \bar{\omega}_{30}^2 = \bar{\omega}_{31}^2 = \bar{\omega}_{32}^2 = \bar{\omega}_{33}^2 = \frac{\left(i\hat{r}^2 - 29164iv^2 + 29038iv - 11197i\right)\sqrt{3}}{12\hat{r}(-1+2v)(1+v)}$$

$$+ \frac{-29164v^2 + (29038 + 328\hat{r})v - \hat{r}^2 - 290\hat{r} - 11197}{12\hat{r}(-1+2v)(1+v)}$$

$$\bar{\omega}_{34}^2 = \bar{\omega}_{35}^2 = \bar{\omega}_{36}^2 = \bar{\omega}_{37}^2 = \bar{\omega}_{38}^2 = \bar{\omega}_{39}^2 = \bar{\omega}_{40}^2 = \bar{\omega}_{41}^2 = \bar{\omega}_{42}^2 = 5.4\bar{\omega}^{*2}$$

$$\bar{\omega}_{43}^2 = \bar{\omega}_{44}^2 = \bar{\omega}_{45}^2 = 6.3\bar{\omega}^{*2}$$

$$\bar{\omega}_{46}^2 = \bar{\omega}_{47}^2 = \bar{\omega}_{48}^2 = \frac{7\left(8v - 9 - \sqrt{144v^2 - 104v + 41}\right)}{4\left(-1 + v + 2v^2\right)}$$

$$\bar{\omega}_{49}^2 = \bar{\omega}_{50}^2 = \bar{\omega}_{51}^2 = \bar{\omega}_{52}^2 = \bar{\omega}_{53}^2 = \frac{\left(-i\hat{r}^2 + 29164iv^2 - 29038iv + 11197i\right)\sqrt{3}}{12\hat{r}(-1+2v)(1+v)}$$

$$+ \frac{-29164v^2 + (29038 + 328\hat{r})v - \hat{r}^2 - 290\hat{r} - 11197}{12\hat{r}(-1+2v)(1+v)}$$

$$\bar{\omega}_{54}^2 = \frac{49v - 77 - \sqrt{3661v^2 - 7546v + 4669}}{2(1+v)(-1+2v)}$$

$$\hat{r} = \sqrt[3]{-7881582v^2 + 4356512v^3 + 4602291v - 1125361 - 27\tilde{r} + 54\tilde{r}v}$$

$$\tilde{r} = \sqrt{-1997914800v^4 - 138713064v^3 + 317906925v^2 + 18932004v - 188425188}$$

(38)

Fig. 2 shows the dependence of the squares of natural frequencies on poison's ratio - $v$, for the third order solution. The first order frequencies, $\bar{\omega}^{*2}$ and $\bar{\omega}_3^2$, are illustrated by the gold lines. It is evident that the first order approximation is not able to capture the lowest frequencies.

Moreover, the third order solution demonstrates that different modes may be of greater engineering importance depending on Poisson's ratio.

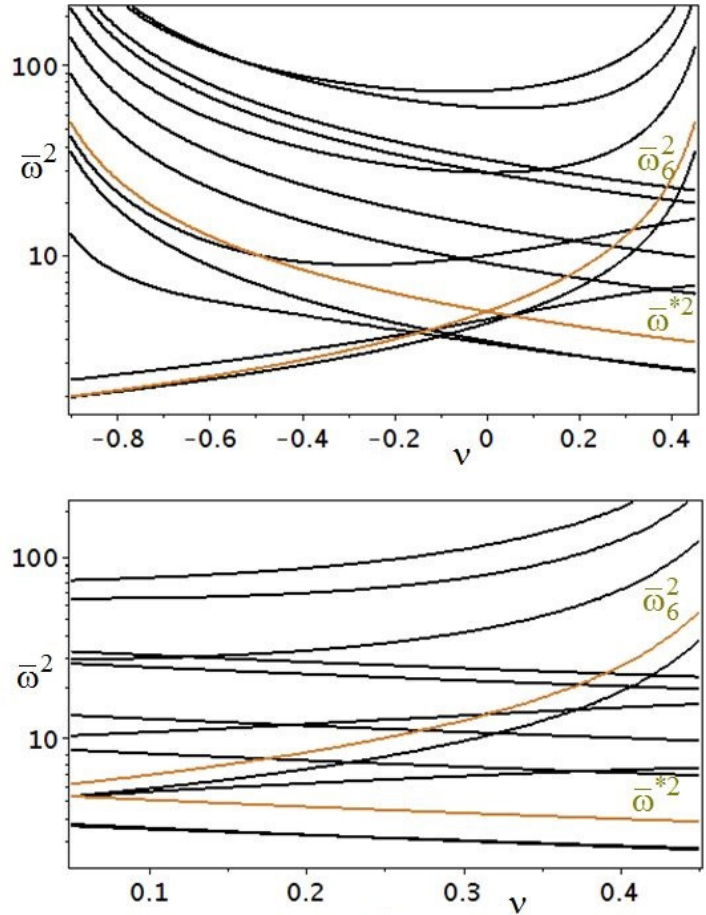

Figure 2: Third order solution showing 54 square non-trivial natural frequencies of a free vibrating sphere as a function of Poisson's ratio. Also, frequencies $\bar{\omega}^{*2}$ and $\bar{\omega}_3^2$, which correspond to the first order approximation, are illustrated in gold. The lower plot is identical to the upper, but focuses on the engineering range of $\nu \in [0.1, 0.4]$.

Next, we examine the accuracy of the approximated solutions. To this end, we compare between the 24 lowest frequencies (excluding the six trivial modes associated with rigid body motion) calculated by first, second, third and fourth order approximations to the results of a finite elements method (FEM). The finite elements analysis was performed with the commercial software ABAQUS™ 6.10, using standard 10 node full integration tetrahedral element *C3D10*, and for the following typical parameters $E = 210 \, [GPa]$, $\rho_0 = 7800 \, [Kg/m^3]$, $\nu = 0.3$, $H = 1 \, [m]$. However, Table 1 consists of $\bar{\omega}$ values for convenience.

Converges of the finite element analysis was validated by increasing the number of elements and examining the tenth lowest natural frequency. FEM results provided in Table 1 were calculated using 35349 elements, and expected to have no more than 0.001% inaccuracy.

| Mode # | FEM | O(1) | O(2) | O(3) | O(4) | Err % O(1) | Err % O(2) | Err % O(3) | Err % O(4) |
|---|---|---|---|---|---|---|---|---|---|
| 1 | 1.551 | 1.961 | 1.641 | 1.641 | 1.551 | 26.43 | 5.78 | 5.78 | 0.02 |
| 2 | 1.551 | 1.961 | 1.641 | 1.641 | 1.551 | 26.43 | 5.78 | 5.78 | 0.02 |
| 3 | 1.551 | 1.961 | 1.641 | 1.641 | 1.551 | 26.43 | 5.78 | 5.78 | 0.02 |
| 4 | 1.551 | 1.961 | 1.641 | 1.641 | 1.551 | 26.43 | 5.78 | 5.78 | 0.02 |
| 5 | 1.551 | 1.961 | 1.641 | 1.641 | 1.551 | 26.43 | 5.78 | 5.78 | 0.02 |
| 6 | 1.641 | 3.536 | 1.961 | 1.647 | 1.647 | 115.43 | 19.50 | 0.35 | 0.35 |
| 7 | 1.641 | --- | 1.961 | 1.647 | 1.647 | --- | 19.50 | 0.35 | 0.35 |
| 8 | 1.641 | --- | 1.961 | 1.647 | 1.647 | --- | 19.50 | 0.35 | 0.35 |
| 9 | 1.641 | --- | 1.961 | 1.647 | 1.647 | --- | 19.50 | 0.35 | 0.35 |
| 10 | 1.641 | --- | 1.961 | 1.647 | 1.647 | --- | 19.50 | 0.35 | 0.35 |
| 11 | 2.189 | --- | 2.481 | 2.481 | 2.193 | --- | 13.33 | 13.33 | 0.16 |
| 12 | 2.189 | --- | 2.481 | 2.481 | 2.193 | --- | 13.33 | 13.33 | 0.16 |
| 13 | 2.189 | --- | 2.481 | 2.481 | 2.193 | --- | 13.33 | 13.33 | 0.16 |
| 14 | 2.397 | --- | 3.282 | 2.631 | 2.476 | --- | 36.92 | 9.78 | 3.29 |
| 15 | 2.397 | --- | 3.282 | 2.631 | 2.476 | --- | 36.92 | 9.78 | 3.29 |
| 16 | 2.397 | --- | 3.282 | 2.631 | 2.476 | --- | 36.92 | 9.78 | 3.29 |
| 17 | 2.397 | --- | 3.282 | 2.631 | 2.476 | --- | 36.92 | 9.78 | 3.29 |
| 18 | 2.397 | --- | 3.282 | 2.631 | 2.476 | --- | 36.92 | 9.78 | 3.29 |
| 19 | 2.397 | --- | 3.282 | 2.631 | 2.476 | --- | 36.92 | 9.78 | 3.29 |
| 20 | 2.397 | --- | 3.282 | 2.631 | 2.476 | --- | 36.92 | 9.78 | 3.29 |
| 21 | 2.442 | --- | 3.536 | 3.102 | 2.631 | --- | 44.8 | 27.02 | 7.7 |
| 22 | 2.442 | --- | 6.186 | 3.282 | 2.631 | --- | 153.3 | 34.39 | 7.7 |
| 23 | 2.442 | --- | 6.186 | 3.282 | 2.631 | --- | 153.3 | 34.39 | 7.7 |
| 24 | 2.442 | --- | 6.186 | 3.282 | 2.631 | --- | 153.3 | 34.39 | 7.7 |

Table 1: Comparison between the 24 lowest natural frequencies calculated by $1^{st}$, $2^{nd}$, $3^{nd}$, $4^{th}$ order approximations and by finite elements method (FEM). Error is calculated with respect to FEM results.

The results of the comparison between the FEM and our approximated solution are outlined in Table 1. It is evident that higher order approximations enhance the accuracy of the solution, as expected. In particular, the fourth order approximation - O(4), provides excellent predictions with less than 8% error. Besides the improved accuracy, the difference between the first order and second order approximations is even more fundamental, since the first order approximation is not able to capture the lowest frequencies.

The characteristic polynomial associated with O(4) is of degree $n^{dof} = 105$. Several roots of this polynomial can be derived analytically, while others need to be found numerically. One of the frequencies which can be obtained analytically is:

$$\bar{\omega}^2_{fundamental} = \frac{3.128887860}{1+\nu} \tag{39}$$

Importantly, we have found that this frequency is the lowest among all O(4) frequencies in the engineering range of $\nu > 0$. In addition, we have compared between (39) and the fundamental (lowest) frequency obtained by FEM. The maximum difference between the two did not exceed 0.02%. This result is of extreme importance. We have provided, for the first time, a closed form expression for the *fundamental frequency* of a free vibrating elastic sphere. This expression provides high accuracy for the entire range of practical materials having positive Poisson's ratio.

# References

[1] O. C. Zienkiewicz and R. L. Taylor, *The finite element method for solid and structural mechanics*: Butterworth-Heinemann, 2005.

[2] J. Hutchinson and S. Zillmer, "Vibration of a free rectangular parallelepiped," *Journal of Applied Mechanics,* vol. 50, p. 123, 1983.

[3] J. Fromme and A. Leissa, "Free vibration of the rectangular parallelepiped," *The Journal of the Acoustical Society of America,* vol. 48, p. 290, 1970.

[4] K. Liew, K. Hung, and M. Lim, "Free vibration studies on stress-free three-dimensional elastic solids," *Journal of Applied Mechanics,* vol. 62, pp. 159-165, 1995.

[5] A. Leissa and Z. d. Zhang, "On the three-dimensional vibrations of the cantilevered rectangular parallelepiped," *The Journal of the Acoustical Society of America,* vol. 73, p. 2013, 1983.

[6] C. Lim, "Three-dimensional vibration analysis of a cantilevered parallelepiped: exact and approximate solutions," *The Journal of the Acoustical Society of America,* vol. 106, p. 3375, 1999.

[7] D. Zhou and O. McGee III, "On the three-dimensional vibrations of elastic prisms with skew cross-section," *Meccanica,* pp. 1-24, 2013.

[8] J. So and A. Leissa, "Free vibrations of thick hollow circular cylinders from three-dimensional analysis," *Journal of vibration and acoustics,* vol. 119, pp. 89-95, 1997.

[9] O. McGee III and J. Kim, "Three-dimensional vibrations of cylindrical elastic solids with V-notches and sharp radial cracks," *Journal of Sound and Vibration,* vol. 329, pp. 457-484, 2010.

[10] J. Hutchinson, "Transverse vibrations of beams, exact versus approximate solutions," *Journal of Applied Mechanics,* vol. 48, p. 923, 1981.

[11] K. Liew, Y. Xiang, and S. Kitipornchai, "Transverse vibration of thick rectangular plates—I. Comprehensive sets of boundary conditions," *Computers & structures,* vol. 49, pp. 1-29, 1993.

[12] K. Liew, K. Hung, and M. Lim, "Three-dimensional elasticity solutions to vibration of cantilevered skewed trapezoids," *AIAA journal,* vol. 32, pp. 2080-2089, 1994.

[13] A. Leissa, J. Lee, and A. Wang, "Vibrations of cantilevered shallow cylindrical shells of rectangular planform," *Journal of Sound and Vibration,* vol. 78, pp. 311-328, 1981.

[14] K. Liew, K. Hung, and M. Lim, "Vibration characteristics of simply supported thick skew plates in three-dimensional setting," *Journal of Applied Mechanics,* vol. 62, pp. 880-886, 1995.

[15] K. Liew and K. Hung, "Three-dimensional vibratory characteristics of solid cylinders and some remarks on simplified beam theories," *International Journal of Solids and Structures,* vol. 32, pp. 3499-3513, 1995.

[16] O. McGee and G. Giaimo, "Three-dimensional vibrations of cantilevered right triangular plates," *Journal of Sound and Vibration,* vol. 159, pp. 279-293, 1992.

[17] M. B. Rubin, "Free-Vibration of a Rectangular Parallelepiped Using the Theory of a Cosserat Point," *Journal of Applied Mechanics-Transactions of the Asme,* vol. 53, pp. 45-50, Mar 1986.

[18] J. M. Solberg and P. Papadopoulos, "Impact of an elastic pseudo-rigid body on a rigid foundation," *International Journal of Engineering Science,* vol. 38, pp. 589-603, Apr 2000.

[19] P. Naghdi, "The theory of plates and shells," *Handbuch der Physik, vol. VIa/2,* pp. 425-640, 1972.

[20] M. Rubin, Author, A. Cardon, and Reviewer, "Cosserat Theories: Shells, Rods and Points. Solid Mechanics and its Applications, Vol 79," *Applied Mechanics Reviews,* vol. 55, pp. B109-B110, 2002.

[21] S. Antman, *Nonlinear problems of elasticity* vol. 107: Springer, 2005.

[22] J. Casey, "The ideal pseudo-rigid continuum," *Proceedings of the Royal Society a-Mathematical Physical and Engineering Sciences,* vol. 462, pp. 3185-3195, Oct 8 2006.

[23] M. Rubin, "On the numerical solution of spherically symmetric problems using the theory of a Cosserat surface," *International Journal of Solids and Structures,* vol. 23, pp. 769-784, 1987.

[24] M. Jabareen, E. Hanukah, and M. Rubin, "A ten node tetrahedral Cosserat Point Element (CPE) for nonlinear isotropic elastic materials," *Computational Mechanics,* pp. 1-29, 2012.

[25] M. Jabareen and M. B. Rubin, "A Generalized Cosserat Point Element (Cpe) for Isotropic Nonlinear Elastic Materials Including Irregular 3-D Brick and Thin Structures," *Journal of Mechanics of Materials and Structures,* vol. 3, pp. 1465-1498, Oct 2008.

[26] M. Jabareen and M. B. Rubin, "A Cosserat Point Element (CPE) for the Numerical Solution of Problems in Finite Elasticity," in *Mechanics of Generalized Continua*, ed: Springer, 2010, pp. 263-268.

[27] M. Jabareen and M. Rubin, "Failures of the three-dimensional patch test for large elastic deformations," *International Journal for Numerical Methods in Biomedical Engineering,* vol. 26, pp. 1618-1624, 2010.

[28] P. Papadopoulos, "On a class of higher-order pseudo-rigid bodies," *Mathematics and Mechanics of Solids,* vol. 6, pp. 631-640, Dec 2001.

[29] D. J. Steigmann, "On pseudo-rigid bodies," *Proceedings of the Royal Society a-Mathematical Physical and Engineering Sciences,* vol. 462, pp. 559-565, Feb 8 2006.

[30] H. Cohen and R. G. Muncaster, *Theory of Pseudo-rigid Bodies.* Berlin: Springer, 1984b.

[31] J. Slawianowski, "The Mechanics of the Homogeneously-Deformable Body. Dynamical Models with High Symmetries," *ZAMM-Journal of Applied Mathematics and Mechanics/Zeitschrift für Angewandte Mathematik und Mechanik,* vol. 62, pp. 229-240, 1982.

[32] R. Muncaster, "Invariant manifolds in mechanics II: Zero-dimensional elastic bodies with directors," *Archive for Rational Mechanics and Analysis,* vol. 84, pp. 375-392, 1984.

[33] J. M. Solberg and P. Papadopoulos, "A simple finite element-based framework for the analysis of elastic pseudo-rigid bodies," *International journal for numerical methods in engineering,* vol. 45, pp. 1297-1314, Jul 30 1999.

[34] O. Zienkiewicz and R. Taylor, *The Finite Element Method*, 6 ed. London U.K.: McGraw-Hill, 2005.

[35] H. Cohen and Q. X. Sun, "Plane Motions of Elastic Pseudo-Rigid Pendulums," *Solid Mechanics Archives,* vol. 13, pp. 147-176, 1988.

[36] H. Cohen and G. P. Macsithigh, "Impulsive Motions of Elastic Pseudo-Rigid Bodies," *Journal of Applied Mechanics-Transactions of the Asme,* vol. 58, pp. 1042-1048, Dec 1991.

[37] E. Hanukah and B. Goldshtein, "A structural theory for a 3D isotropic linear-elastic finite body," *arXiv preprint arXiv:1207.6767,* 2012.